\documentclass[sigconf]{acmart}
\pdfoutput=1
\AtBeginDocument{%
  \providecommand\BibTeX{{%
    \normalfont B\kern-0.5em{\scshape i\kern-0.25em b}\kern-0.8em\TeX}}}

\copyrightyear{2024}
\acmYear{2024}
\setcopyright{acmlicensed}\acmConference[EASE 2024]{28th International Conference on Evaluation and Assessment in Software Engineering}{June 18--21, 2024}{Salerno, Italy}
\acmBooktitle{28th International Conference on Evaluation and Assessment in Software Engineering (EASE 2024), June 18--21, 2024, Salerno, Italy}
\acmDOI{10.1145/3661167.3661214}

\acmConference[EASE 2024]{The 28th International Conference on Evaluation and Assessment in Software Engineering}{18–21 June, 2024}{Salerno, Italy}
\acmPrice{15.00}
\acmISBN{979-8-4007-1701-7/24/06}

\usepackage{graphicx}
\usepackage{tikz,lipsum}
\usepackage{xcolor}
\usepackage{soul}

\definecolor{gray10}{gray}{.9}
\definecolor{arsenic}{rgb}{0.23, 0.27, 0.29}
\usepackage{color}

\RequirePackage{fontawesome}

\definecolor{gray50}{gray}{.5}
\definecolor{gray40}{gray}{.6}
\definecolor{gray30}{gray}{.7}
\definecolor{gray20}{gray}{.8}
\definecolor{gray10}{gray}{.9}
\definecolor{gray05}{gray}{.95}

\newlength\Linewidth
\def\findlength{\setlength\Linewidth\linewidth
  \addtolength\Linewidth{-4\fboxrule}
  \addtolength\Linewidth{-3\fboxsep}
}
\newenvironment{examplebox}{\par\begingroup
  \setlength{\fboxsep}{5pt}\findlength
  \setbox0=\vbox\bgroup\noindent
  \hsize=0.95\linewidth
  \begin{minipage}{0.95\linewidth}\normalsize}
  {\end{minipage}\egroup
  \textcolor{gray20}{\fboxsep1.5pt\fbox
    {\fboxsep5pt\colorbox{gray05}{\normalcolor\box0}}}
  \endgroup\par\noindent
  \normalcolor\ignorespacesafterend}

\usepackage{balance}
\usepackage{comment}

\begin{document}

\title{Trustworthy AI in practice: an analysis of practitioners' needs and challenges}

\author{Maria Teresa Baldassarre}
\orcid{0000-0001-8589-2850}
\affiliation{%
  \institution{University of Bari "A. Moro"}
  \city{Bari}
  \country{Italy}}
\email{mariateresa.baldassarre@uniba.it}

\author{Domenico Gigante}
\orcid{0000-0003-3589-6970}
\affiliation{%
  \institution{Ser\&Practices Srl}
  \city{Bari}
  \country{Italy}}
\email{d.gigante@serandp.com}
\authornote{Authors are listed in alphabetical order. Corresponding authors: Domenico Gigante (domenico.gigante1@uniba.it) and Azzurra Ragone (azzurra.ragone@uniba.it).}

\author{Marcos Kalinowski}
\orcid{0000-0003-1445-3425}
\affiliation{%
  \institution{Pontifical Catholic University of Rio de Janeiro (PUC-Rio)}
  \city{Rio de Janeiro}
  \country{Brazil}}
\email{kalinowski@inf.puc-rio.br}


\author{Azzurra Ragone}
\orcid{0000-0002-3537-7663}
\authornotemark[1]
\affiliation{%
  \institution{University of Bari "A. Moro"}
  \city{Bari}
  \country{Italy}}
\email{azzurra.ragone@uniba.it}

\author{Sara Tibidò}
\affiliation{%
  \institution{Scuola IMT Alti Studi Lucca}
  \city{Bari}
  \country{Italy}
}
\email{sara.tibido@imtlucca.it}
\renewcommand{\shortauthors}{Baldassarre et al.}

\begin{abstract}
Recently, there has been growing attention on behalf of both academic and practice communities towards the ability of Artificial Intelligence (AI) systems to operate responsibly and ethically. 
As a result, a plethora of frameworks and guidelines have appeared to support practitioners in implementing Trustworthy AI applications (TAI).
However, little research has been done to investigate whether such frameworks are being used and how.
In this work, we study the vision AI practitioners have on TAI principles, how they address them, and what they would like to have -- in terms of tools, knowledge, or guidelines -- when they attempt to incorporate such principles into the systems they develop. 
Through a survey and semi-structured interviews, we systematically investigated practitioners' challenges and needs in developing TAI systems. 
Based on these practical findings, we highlight 
recommendations to help AI practitioners develop Trustworthy AI applications. 
\end{abstract}



\maketitle

\section{Introduction}
\label{sec:intro}

Artificial intelligence (AI) systems increasingly exert an extensive impact on various facets of our existence, encompassing the realm of healthcare and the quality of education we receive \cite{Bosch2016DetectingSE, Esteva2017DermatologistlevelCO, Holstein2018StudentLB}, the determination of which news articles or social media posts we encounter \cite{Alvarado2018TowardsAE, Bucher2017, Rader2015UnderstandingUB}, the allocation of employment opportunities \cite{HireVuecom, pymetricsai}, the detention decisions \cite{Chouldechova2016FairPW}, and the intensification of policing efforts in some areas \cite{Lum2016, 10.1145/3173574.3174014}, just to name a few. 
With this expansion, the risk of AI increasing social inequities has generated escalating attention across several communities, including the media. Indeed it is common to observe reports in mainstream media of systemic dangerous behaviors observed in widely used AI systems, such as a smart algorithm guiding assistance for tens of millions of people biased against dark-skinned patients\footnote{\url{https://www.nature.com/articles/d41586-019-03228-6}}, or an AI chatbot suspended for making homophobic slurs and leaking user information\footnote{\url{https://www.ic3.gov/Media/News/2021/210310-2.pdf}}.
These risks are even more pronounced with the recent advent of Generative AI and the impact these systems have on various societal aspects \cite{10.1145/3582515.3609555}.

In this context, the concept of Trustworthy Artificial Intelligence (TAI) has been defined: "\textit{Trustworthy AI has three components, which should be met throughout the system's entire life cycle: (1) it should be lawful, complying with all applicable laws and regulations (2) it should be ethical, ensuring adherence to ethical principles and values and (3) it should be robust, both from a technical and social perspective since, even with good intentions, AI systems can cause unintentional harm. Each component in itself is necessary but not sufficient for the achievement of Trustworthy AI}" \cite{AIHLEG_ethics_guidelines}. 

Several public and private organizations have responded to these societal fears by developing different kinds of resources: ethical requirements, principles, guidelines, best practices, tools, and frameworks \cite{10.1145/3593434.3593478, Pardo2011}.
As the field progresses, integrated toolkits are being developed with the objective of rendering these methods more broadly accessible and usable (\textit{e.g.}, Enisa's Machine Learning Security \cite{enisaml}, Aequitas \cite{aequitas} or Google's What-if-Tool \cite{whatiftool}). 

Despite growth in the development and dissemination of toolkits, there has been little research investigating how AI practitioners actually use them in practice throughout the entire Software Development Life Cycle (SDLC). A few studies have already explored practitioners' perceptions and desires around open-source toolkits but are focused only on the Fairness aspect of TAI \cite{10.1145/3411764.3445261, Richardson2021TowardsFI}. 
However, neither of these two works investigated what professionals need during the entire SDLC.

Our research builds on the results of our previous mapping study \cite{10.1145/3593434.3593478} and aims to investigate how AI professionals deal with TAI issues on a day-to-day basis.
To better understand practitioners' needs 
we distributed a \textbf{survey} and conducted semi-structured one-on-one \textbf{interviews}, collecting data from a total of \textbf{34 practitioners} 
employed in companies of different sizes. 


As the main contribution, this work intends to deeply investigate practitioners' views, needs, and challenges in developing TAI systems throughout the entire SDLC.
The novel contribution can be summarized as follows:
\begin{itemize}
    \item We have analyzed the \textbf{existing procedures} that development teams adopt when implementing Trustworthiness in AI;
    \item We have investigated the \textbf{impediments} encountered in the attempt to implement Trustworthiness in AI;
    \item We have identified a range of practitioners'  \textbf{needs} that academic and industrial research should seek to address.
\end{itemize}
Our study has identified a range of practitioner needs that have thus far been overlooked in the literature. 
For example, 
the majority of our interviewees report a lack of tools for the late stages of the SDLC, as well as of knowledge bases and practical guidelines with suggestions on implementing TAI across the entire SDLC. 

The paper is organized as follows. Section \ref{sec:background} provides some background definitions. 
Section \ref{sec:Study Design} describes the methodology we adopted to conduct this study, including research questions and how we collected, extracted, and analyzed the data. Section \ref{sec:results} presents the quantitative and qualitative results together with some preliminary findings. 
Section \ref{sec:discussion}  discusses the results, and practical implications and recommendations for the AI industry and research community.
Section \ref{sec:threats} addresses threats to validity, followed by the conclusion in Section \ref{sec:conclusions}.

\section{Background}
\label{sec:background}


\subsection{AI Principles proliferation}
\label{subsec:ai_principles}

National and global entities have established specialized expert committees in the field of Artificial Intelligence (AI) to address the associated risks stemming from AI development. These committees often have the mandate of formulating policy documents. Prominent examples of such organizations include the High-Level Expert Group on Artificial Intelligence initiated by the European Commission \cite{AIHLEG},
the UNESCO Ad Hoc Expert Group (AHEG) tasked with the Recommendation on the Ethics of Artificial Intelligence \cite{AHEG},
the Advisory Council on the Ethical Use of Artificial Intelligence and Data in Singapore \cite{singapore_council},
the NASA Artificial Intelligence Group \cite{nasa_ai_group}
and the UK AI Council \cite{uk_ai_council},
among others.

These committees bear the crucial role of generating comprehensive reports and guidelines about Trustworthy AI (TAI). A parallel endeavour is observable within the commercial landscape, particularly among enterprises heavily reliant on AI technologies. Corporations such as Sony\footnote{\url{https://www.sony.com/en/SonyInfo/sony_ai/responsible_ai.html}} and Meta\footnote{\url{https://ai.facebook.com/blog/facebooks-five-pillars-of-responsible-ai/}} have made their AI policies and principles publicly accessible. Concurrently, professional organizations and non-profit entities, such as UNI Global Union\footnote{\url{http://www.thefutureworldofwork.org/media/35420/uni_ethical_ai.pdf}} and the Internet Society\footnote{\url{https://www.internetsociety.org/resources/doc/2017/artificial-intelligence-and-machine-learning-policy-paper/}}, have issued statements and recommendations.

The substantial efforts of this diverse spectrum of stakeholders in crafting TAI principles and policies not only underscore the imperative need for ethical guidance but also exemplify their vested interest in shaping AI ethics to align with their specific objectives \cite{Greene2019BetterNC}. It is noteworthy that the private sector's engagement in the gap of AI ethics has undergone a thorough check, with contentions suggesting that high-level soft policies may be employed to either transform a social issue into a purely technical one \cite{Greene2019BetterNC} or to potentially circumvent regulatory measures \cite{bayamlioglu_being_2018, Jobin2019}.

Nevertheless, a set of research endeavors has brought attention to the divergent nature of these proposals, giving rise to a complex challenge often referred to as \textit{"principle proliferation"}\textcolor{black}{\footnote{It is ''\textit{the proliferation of soft-law efforts}''. Jobin’s analysis \cite{Jobin2019} shows the emergence of an apparent cross-stakeholder convergence on promoting the ethical principles [...] [but] unclarity remains as to which ethical principles should be prioritized, how conflicts between ethical principles should be resolved, who should enforce ethical oversight on AI and how researchers and institutions can comply with the resulting guidelines.}} \cite{Jobin2019}. Consequently, efforts have been undertaken to address this challenge. For instance, Jobin et al. \cite{Jobin2019} conducted a comprehensive study, that culminated in the identification of a global convergence around five ethical principles: \textit{transparency}, \textit{justice and fairness}, \textit{non-maleficence}, \textit{responsibility}, and \textit{privacy}. Jobin et al. \cite{Jobin2019} also observed that, while no single document they reviewed encompassed all of these ethical principles, these five principles were mentioned in over half of the sources examined. Furthermore, their detailed thematic analysis unveiled significant semantic and conceptual variations in the interpretation of these principles and the specific recommendations or areas of concern derived from each one.

\subsection{Trustworthy AI principles definitions}
\label{subsec:chosen_definition}



As set out in Section \ref{subsec:ai_principles}, a notable degree of ambiguity and subtlety exists in demarcating the principles that predominantly characterize Trustworthy AI (TAI). Notably, TAI is sometimes used interchangeably with Responsible or Ethical AI. In our investigation, we confront the challenge of \textit{principle proliferation} by choosing to focus on a specific subset of principles that characterize TAI. Specifically, we concentrate on the most recurrent four principles identified by Jobin et al. \cite{Jobin2019}, while opting to exclude the principle of \textit{responsibility} due to its infrequent occurrence and lack of a clear, universally accepted definition.

Furthermore, in this work, we have decided to adopt the definitions put forth by the High-Level Expert Group on Artificial Intelligence (AIHLEG)
--- an entity established by the European Commission --- explained in their "\textit{Ethics guidelines for trustworthy AI}" \cite{AIHLEG_ethics_guidelines}. 

Given these premises, we mapped the four selected principles identified by Jobin et al. \cite{Jobin2019}, with the formal definition delineated by the AIHLEG \cite{AIHLEG_ethics_guidelines}. The mapping has been carried out based on the contents of the definitions and not merely on the nomenclatures, as, in most cases, they differ.
For the sake of simplicity, we have shortened and labeled each TAI principle as follows: \textbf{Transparency}, \textbf{Fairness}, \textbf{Security}, and \textbf{Privacy}. We will use these labels throughout the paper. 
\textcolor{black}{Definitions of the terms adopted for each principle are provided in our online appendix
\cite{online_appendix}}.

\section{Method}
\label{sec:Study Design}


The goal of our study is to investigate the \textbf{state of the practice} to understand common practices as well as challenges and difficulties encountered by AI practitioners in implementing TAI systems through the entire SDLC.

Since traditional phases of the SDLC do not necessarily map well with the activities required to develop an AI-enabled system, we have extended each phase with one or more activities mentioned by Zhengxin et al. \cite{Zhengxin2023MLOpsSW}. Table \ref{table:sdlc_mlops} graphically shows how each traditional SDLC phase's definition has been extended.

\begin{table}[]
\caption{Activities integrated into the traditional SDLC phases to support AI-enabled systems development.}
\begin{center}
\large
\label{table:sdlc_mlops}
\begin{tabular}{ll}
\hline
\textbf{Traditional SDLC phase} & \textbf{Integrated activity}                                                   \\ \hline
Requirements Elicitation        & Model Requirement                                                              \\ \hline
Design                          & \begin{tabular}[c]{@{}l@{}}Data Collection, \\ Data Preparation\end{tabular}   \\ \hline
Development                     & \begin{tabular}[c]{@{}l@{}}Feature Engineering, \\ Model Training\end{tabular} \\ \hline
Test                            & Model Evaluation                                                               \\ \hline
Deployment                      & Model Deployment                                                               \\ \hline
Monitoring                      & Model Monitoring                                                               \\ \hline
\end{tabular}
\end{center}
\end{table}


Given our main goal, and based on the analysis of the shortcomings and common practices that emerged from the mapping study carried out in our previous work \cite{10.1145/3593434.3593478},
we defined the following Research Questions (RQs):
\begin{itemize}
    \item \textbf{RQ1}: What vision/opinion do practitioners have about untrustworthiness issues and how often do they encounter them?
    \item \textbf{RQ2}: How do practitioners address untrustworthiness issues?
    \item \textbf{RQ3}: What tools/support do practitioners desire to address untrustworthiness problems better?
\end{itemize}


\subsection{Data collection}
\label{subsec:search_strategy}

For our study, we recruited practitioners\footnote{By "AI practitioners" we mean those who work in any role
on a team developing products or services involving AI.} working on AI  products and services through a combination of purposive and snowball sampling \cite{Heckathorn2011CommentSV}. 
To recruit participants for our study, we chose to not widely distribute our survey online, but rather to conduct a brief screening procedure to define the inclusion criteria and target suitable participants for the study. 
The target participants for the survey were AI practitioners involved in the development of AI-based systems, with at least some basic knowledge of TAI principles and/or who had previously addressed TAI in their professional work. 
We started by sending personal emails to contacts within our network, working in industry or academia, explaining the purpose of the study and the inclusion criteria, we asked them to help us recruit other participants by spreading the invitation through their networks. 
Next, we verified that the inclusion criteria were met by asking some specific questions in the demographics section of the survey.

The invitation to participate in the study was sent by email and included an explanation of the study's purpose.  
\textcolor{black}{In order to meet all needs (\textit{e.g.}, restrictions related to tight schedules, time zones, commitments), we gave participants the chance to choose between (a) survey (asynchronous interaction) or (b) semi-structured interviews (synchronous interaction) mode. This allowed us to include a broader amount of subjects and collect a higher number of answers. All participants answered the same set of questions and had the opportunity to add any non-listed practices or suggestions/feedback related to the closed-ended question in the open-text fields (asynchronous mode) whereas, interviewed participants (synchronous) could “discuss their answers out loud” and further elaborate their considerations with the interviewers. The answers were all transcribed to be included later in the thematic analysis.} 
 
\textcolor{black}{Participation was on a voluntary basis and not rewarded by any means.} The survey can be accessed at \cite{survey_link}. Overall, we obtained \textbf{23 answers} for the \textbf{survey} and \textbf{11 participants} attended the \textbf{interview}.




Section \ref{sec:demographics} provides details about participants' demographics and their relevant experience. 
Specific details about their companies and working environment have been abstracted to preserve anonymity. 

\subsection{The survey}
\label{subsec:survey}

The survey contains six main sections \cite{survey_link}: 

1. \textit{Informed consent request}. This page asks the participants to provide their informed consent and explains the purpose of the research, the participants' requirements, confidentiality rules, participation on a voluntary basis, and the time needed to complete the survey.


2. \textit{Preliminary concepts knowledge}. We clarified the semantics and interpretation of each TAI principle for participants by providing a definition for each principle. By listing a definition, we wanted to build a shared understanding of each principle to answer the remaining questions.
In addition, we asked the participants about their vision and previous experience with TAI. 

3. \textit{Practices in preventing untrustworthiness in AI}. We inquired participants about the main strategies they adopt to prevent TAI issues (\textit{e.g.}, balancing the dataset or choosing a specific algorithm).

4. \textit{Practices in discovering untrustworthiness issues in AI}. We asked participants to share their strategies to find possible sources of untrustworthiness (\textit{e.g.}, do auditing tasks, compute metrics, learn from user feedback). 

5. \textit{Practices in addressing trustworthiness issues}. We investigated the different approaches used to address TAI issues (\textit{e.g.}, dataset augmentation, instance weighting). 

6. \textit{Demographics and background information}. Participants were asked about gender, level of education, country, role, years of experience, size of the organization, etc.



All the questions of the survey sections (3), (4), (5), illustrated above, as well as the options for the answers, were inspired from and based on our previous work \cite{10.1145/3593434.3593478}, which set current shortcomings and common practices in literature.

The survey  
was anonymous and did not ask for any directly identifying information. 
Most of the survey questions were closed-answer and mandatory, but there were also optional open-text ones.
Through the latter, we were able to collect further qualitative data as many of the participants provided information on the practices they usually implement. All survey data and raw material can be accessed in the online appendix \cite{online_appendix}.



\subsection{Interview Study Protocol}
\label{subsec:interview_protocol}
Before scheduling the interviews with participants, to understand the challenges and requirements for conducting remotely semi-structured interviews, as well as to help us refine our study protocol, we conducted two pilot interviews. 
These preliminary interviews helped us define the protocol and the setting we applied to the final, larger sample.

All interviews were conducted via Microsoft Teams\footnote{\url{https://www.microsoft.com/it-it/microsoft-teams}}; after asking each participant for their consent, we enabled \textit{Recording \& Transcription} Teams features. Once the interview ended, we proof-checked the transcription in order to correct any misspellings, anonymized any Personal Identifiable Information (PII) and finally deleted the recording.

The interview study consisted of think-aloud semi-structured interviews, each one lasting between 45 and 90 minutes. During the live interview, we periodically asked participants to elaborate on their responses, especially for the open ones. We also encouraged participants to "think aloud" \cite{nu11010132, Someren1994} and discuss the information that was being displayed and how their understanding of the question was developing.

To give a standard structure to each interview, we used the survey as a canvas. 

\subsection{Data Analysis}
\label{subsec:extraction_procedure}

In this step, we extracted all relevant data using quantitative and qualitative data analysis techniques to summarize and interpret the collected data. For quantitative data, we used descriptive statistics \cite{Darren2018ds}, and for qualitative data, we used thematic analysis \cite{Cruzes2011rsftasise}.

We used an \textbf{inductive thematic analysis} approach \cite{Braun2006, Braun2012} to analyze about 11.5 hours of video recordings and their corresponding (automatically generated and manually proof-checked) transcripts. The entire analysis was done through Atlas.ti\footnote{\url{https://atlasti.com/}}. 
\textcolor{black}{Two authors worked independently and used the tool to conduct an open coding of the transcripts for each quotation. Next, they manually reviewed each code and decided which to include/exclude annotating any comments. Once this step was completed, they joined to compare and discuss results. The total number of analyzed quotations was 23. The calculated Cohen’s Kappa \cite{Bakeman_Quera_2011} is 0.259. All the details about the coding procedure and the generated codes are provided in the online appendix \cite{online_appendix}.}

\textcolor{black}{Once finalized, the codes were shared with the entire research team and grouped into higher-level themes concerning the practitioners' knowledge and practices. In Section \ref{sec:results} we discuss the findings identified from these codes and themes, together with implications for future TAI developments.}




\section{Results and findings}
\label{sec:results}
We present findings from our think-aloud interviews study and the survey answers, divided into three main sections, 
\begin{itemize}
    \item Practices in \textbf{preventing} untrustworthiness in AI
    \item Practices in \textbf{discovering} untrustworthiness issues
    \item Practices in \textbf{addressing} untrustworthiness issues
\end{itemize}
Across all three phases, we discovered different nuances of practitioners' needs around TAI issues. We supplement data from the closed-ended questions (quantitative results) with the thematic analysis performed on the answers from all the open-ended questions (qualitative results). 
\textcolor{black}{We performed analysis on the disaggregated data with respect to subgroups such as company size, gender, education, and number of projects deployed. To understand if the differences were statistically significant, we conducted pairwise comparisons using Fisher's exact test coupled with the \textit{Benjamini-Hochberg} \cite{benjamini1995controlling} correction to obtain the adjusted p-values. Since there is no statistical significance in any of the cases except for company size, detailed in Section \ref{subsec:preventing_tai}, the graphs in the paper report the results of the analyses in aggregated form.}

\subsection{Preliminary concepts knowledge}
\label{subsec:preliminary_knowledge}
\textcolor{black}{All discussed tables and graphs, from now on, bring together the answers from both the interviews (11) and the survey (23).}
In the survey section "2. \textit{Preliminary concepts knowledge}", 
we observed that the TAI principle \textcolor{black}{participants have encountered most frequently in their projects} is \textbf{Privacy} (20 answers), followed by \textbf{Transparency} (18 answers) and \textbf{Security} (17 answers), while the least experienced is \textbf{Fairness} (13 answers).
This answer should be further investigated because perhaps sometimes practitioners may not recognize or be aware of the need to address some issues related to these principles.


In addition, Table \ref{table:reasons_care_TAI} shows the reasons participants agreed on concerning why they care about TAI. 
The most agreed reasons were \textit{Avoid violating legal requirements} and \textit{Improve the overall quality}.
While, the least agreed one was \textit{Retain users/avoid losing the activity}, with eight disagreements.

\begin{table}[]
\footnotesize
\caption{Reasons why participants care about TAI principles.}
\label{table:reasons_care_TAI}
\begin{tabular}{lcccc}
\hline
\textbf{Reason}                                                                            & \textbf{\begin{tabular}[c]{@{}c@{}}Disagreement \\ (1-2)\end{tabular}} & \textbf{\begin{tabular}[c]{@{}c@{}}Neutral \\ (3)\end{tabular}} & \textbf{\begin{tabular}[c]{@{}c@{}}Agreement \\ (4-5)\end{tabular}} & \textbf{N/A} \\ \hline
\begin{tabular}[c]{@{}l@{}}Doing something \\ about trustworthiness \\ in AI/ML\end{tabular} & 4                                                                      & 6                                                               & 23                                                                  & 1            \\ \hline
\begin{tabular}[c]{@{}l@{}}Avoid violating \\ legal requirements\end{tabular}                & 2                                                                      & 3                                                               & 27                                                                  & 2            \\ \hline
\begin{tabular}[c]{@{}l@{}}Avoid reputational\\ damages\end{tabular}                         & 4                                                                      & 8                                                               & 21                                                                  & 1            \\ \hline
\begin{tabular}[c]{@{}l@{}}Improve the overall\\ quality\end{tabular}                        & 3                                                                      & 3                                                               & 26                                                                  & 2            \\ \hline
\begin{tabular}[c]{@{}l@{}}Retain users/avoid\\ losing   business\end{tabular}               & 8                                                                      & 4                                                               & 19                                                                  & 3           \\ \hline
\end{tabular}
\end{table}

Other important factors related to the reasons for caring about TAI emerged: i) "\textit{need to solve mission-critical tasks}"; ii) "\textit{[need to provide] models usable in real-world contexts}", and this demonstrated that black box models are not allowed in some specific contexts; iii) "\textit{the robustness of the AI explanations themselves}", which shows consciousness about the fact that all TAI aspects contribute to making the model more robust; and iv) "\textit{[need to] desire to commercially assemble AI systems to improve society}".

\begin{figure}[ht]
\includegraphics[scale=0.42]{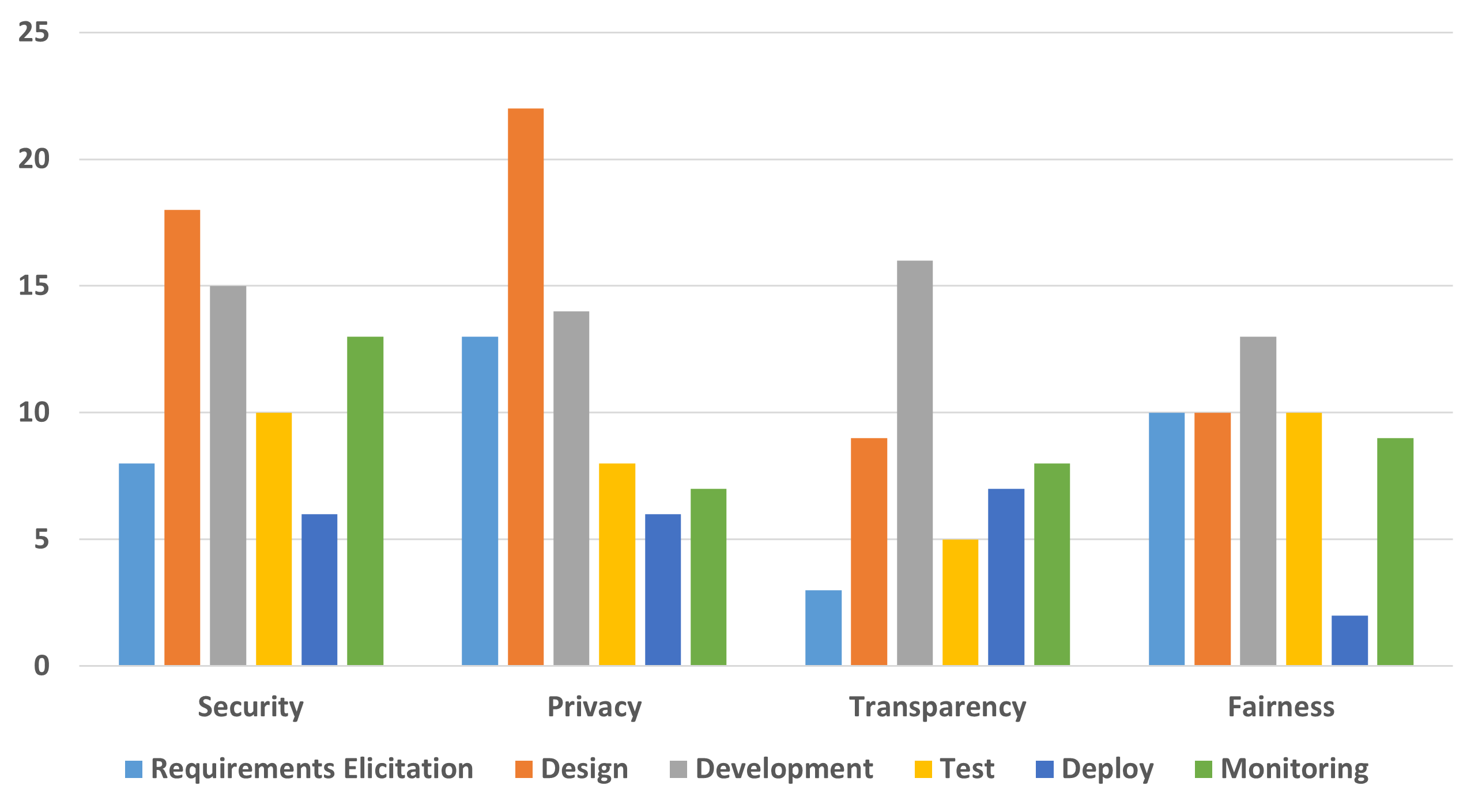}
\caption{TAI principles addressed by SDLC phase.}
\label{fig:tai_issues_addressed_SDLC}
\end{figure}

Finally, Fig. \ref{fig:tai_issues_addressed_SDLC} shows that most of the participants address TAI principles during the \textbf{Design} and \textbf{Development} SDLC phases. In contrast, very few participants reported that they had addressed TAI principles during the \textbf{Requirements Elicitation} and, especially, \textbf{Deploy} phases.



\subsection{Practices in Preventing untrustworthiness in AI}
\label{subsec:preventing_tai}
\begin{figure}[ht]
\includegraphics[scale=0.9]{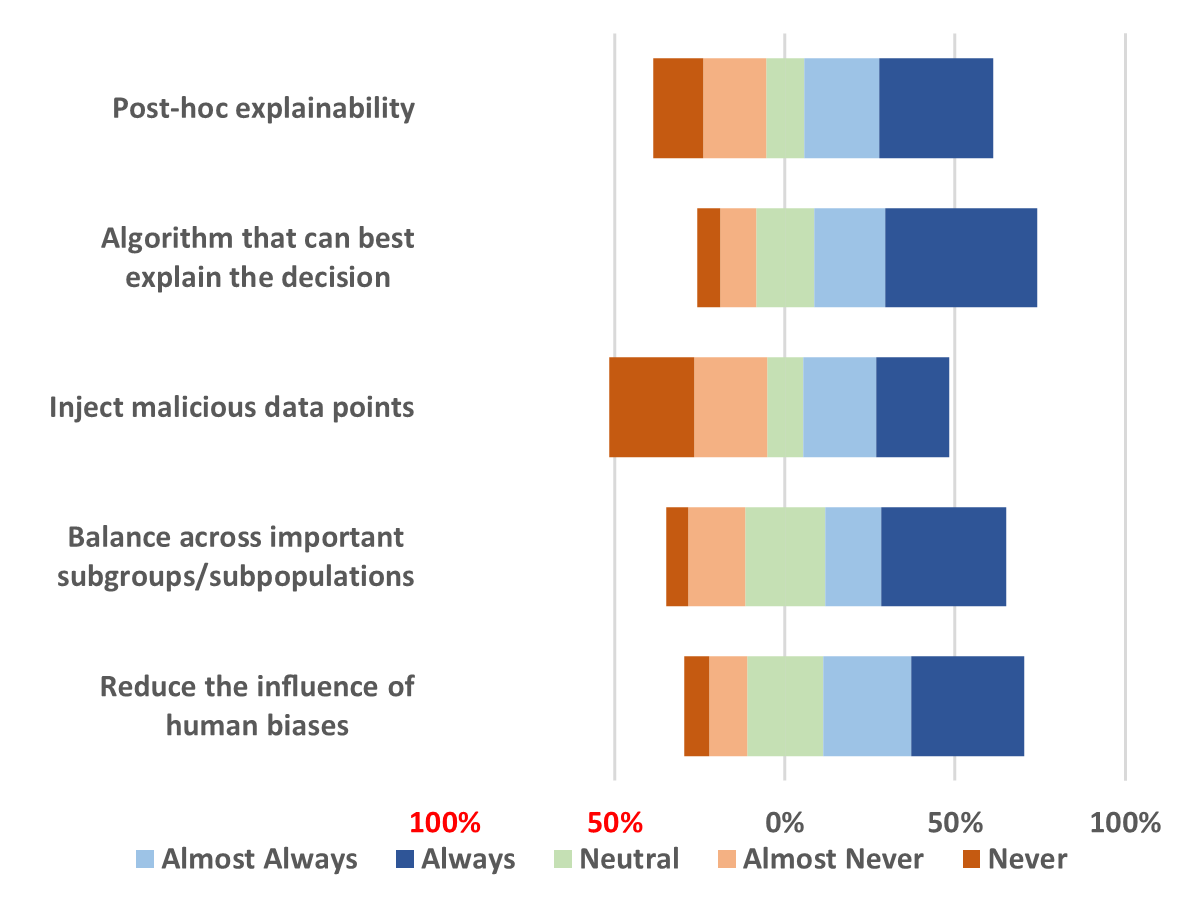}
\caption{Strategies employed to ensure trustworthiness in AI. N/A answers have been removed.}
\label{fig:strategy_prevent_tai_issues}
\end{figure}

In the survey section "3.\textit{Practices in preventing untrustworthiness in AI}", as Fig. \ref{fig:strategy_prevent_tai_issues} shows, the most recurrent strategy employed by our participants to ensure trustworthiness is "\textit{algorithm that can best explain the decision}". On the other hand, the least employed practice appears to be "\textit{inject malicious data points}". 
\textcolor{black}{In analyzing the disaggregated data, we found a statistically significant difference only in the responses related to the strategy “\textit{Algorithm that can best explain the decision}” for the company size subgroup. Indeed, for medium-sized companies, we found more positive responses than for small and large enterprises. For medium ones, no negative answers were given and 63\% of the participants chose "Always", reflecting the wide use of this strategy in medium-sized companies.}



Some participants mentioned other strategies, such as "\textit{[conduct an] in-depth study of the state of the art [prior to start designing the system]}" and "\textit{[use the] post-processing phase [...] to apply human-friendly deterministic rules to check whether a result is in line with the sense of the application domain}".



\begin{figure}[ht]
\includegraphics[scale=0.9]{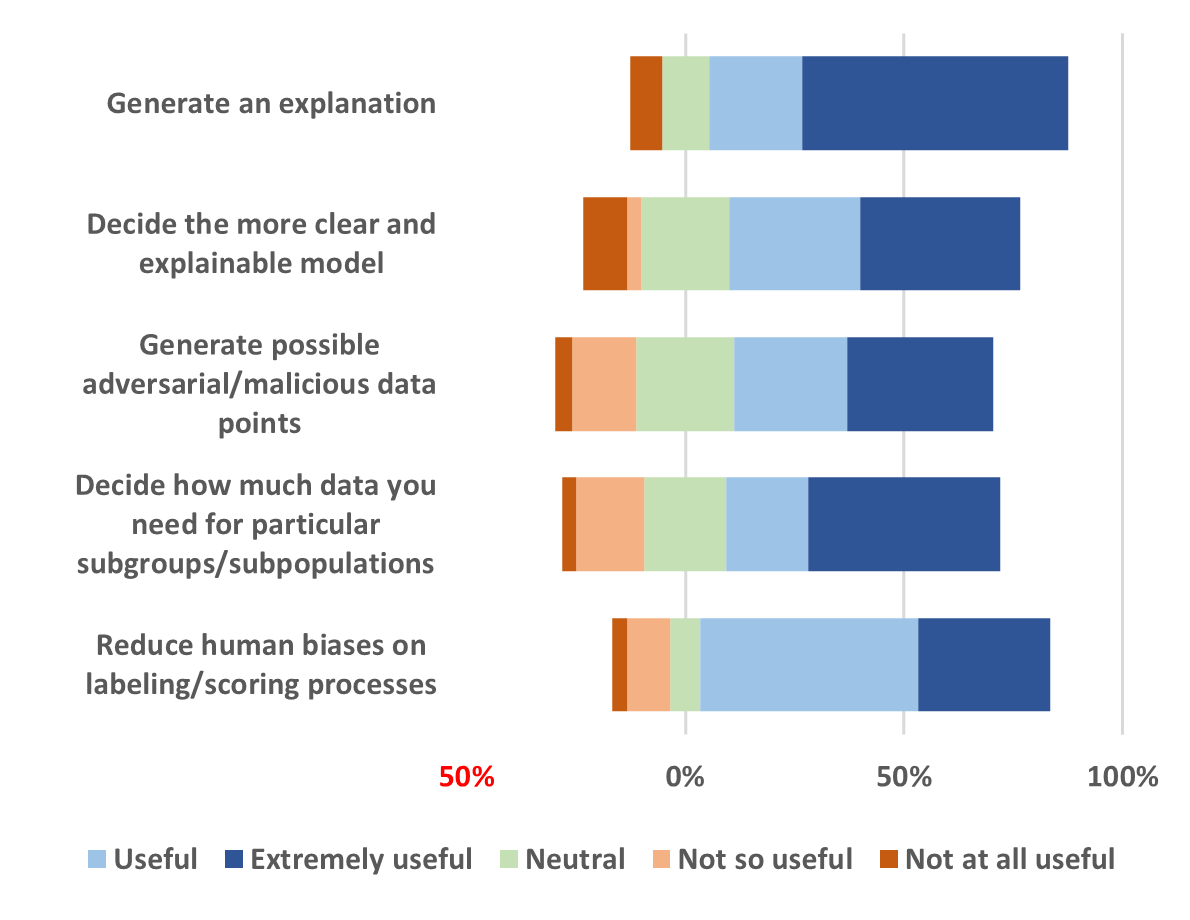}
\caption{Perceived usefulness of hypothetical tools to prevent TAI issues. N/A answers have been removed.}
\label{fig:tool_prevent_tai_issues}
\end{figure}

Additionally, when we asked the participants to rate the utility of various hypothetical tools assuming their team had access to them, the participants rated as the most valuable the tool able to "\textit{[...] generate an explanation of a model after its creation [...]}". On the other hand, they rated as least useful the tool to "\textit{decide how much data you need for particular subgroups/subpopulations}". These results are shown in Fig. \ref{fig:tool_prevent_tai_issues}.



Here too, some important insights emerged, including "\textit{tools to improve software architectures}" and "\textit{[a tool for] referencing the best architecture (dependencies, docker files, instant compute, ...) to perform the task with the lowest possible costs}".

The answers to this section show practitioners are prone to use techniques and tools to prevent trustworthiness issues, focusing mainly on ensuring \textbf{Transparency} (a.k.a. Explainability).

\subsection{Practices in Discovering Untrustworthiness Issues}
\label{subsec:discovering_tai}
In the survey section "4.\textit{Practices in Discovering untrustworthiness in AI}", we investigated which strategies participants mainly employed to discover TAI issues.



\begin{figure}[ht]
\includegraphics[scale=0.43]{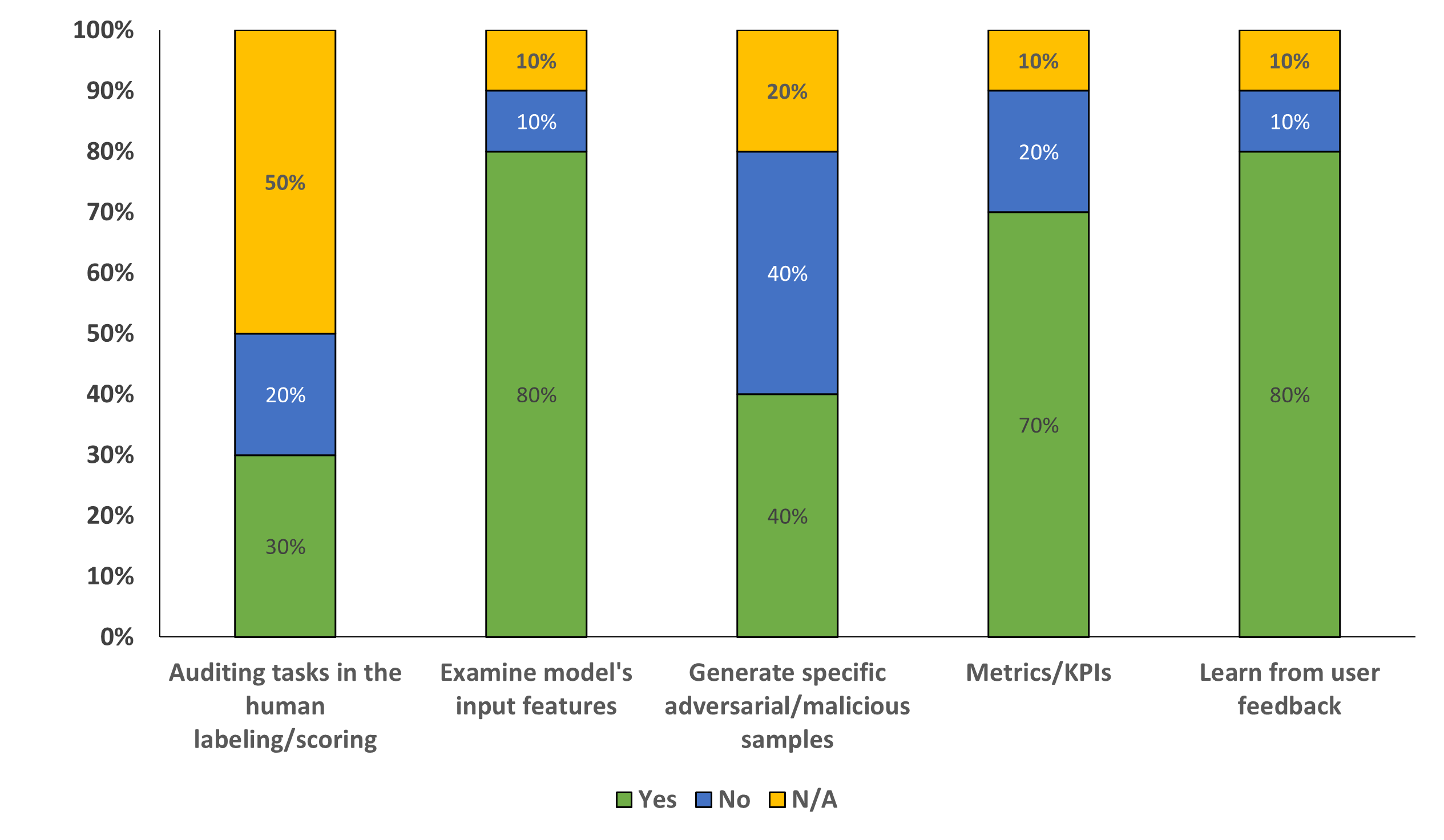}
\caption{Strategies employed to discover untrustworthiness in AI.}
\label{fig:strategies_discover_tai_issues}
\end{figure}

The data shows that the most used strategies are "\textit{Metrics/KPIs}", "\textit{learn from user feedback}", and "\textit{examine AI/ML model's input features}" (see Fig.\ref{fig:strategies_discover_tai_issues}).
\textcolor{black}{Examples of \textit{Metrics/KPIs} related to fairness are, just to cite a few, Demographic Parity, Accuracy, F1-Score. Whereas, \textit{user feedback} is intended as having a feedback form where the users can report misbehavior by the algorithm \cite{gemini_wrong}}.
What stands out is that the strategy less employed by the participants is "\textit{generate specific adversarial/malicious samples}" (4/10 negative answers).


Moreover, when asked for other (not mentioned) strategies they employ, one participant answered "\textit{post processing studies to evaluate possible model 'discriminations'}". As a free thought, another declared "\textit{apply confidence criteria such that it is possible to measure how often the model fails to respond reliably}".

What emerges in this section is that participants employ both qualitative (\textit{e.g.}, \textit{auditing tasks in the human labeling/scoring process}) and quantitative (\textit{e.g.}, \textit{metrics and KPIs}) strategies.

\subsection{Practices in Addressing Untrustworthiness Issues}
Regarding survey section "5.\textit{Practices in Addressing Untrustworthiness in AI}", participants reported that after finding a TAI issue only in 35\% of the cases (12/34) the team addressed it directly, while 15\% (5/34) of the participants stated that it was not addressed by them, but handled by a third party.
Worth noting is the fact that in 50\% (17/34) of the cases participants reported that they did not fix the issue after finding it. 
The reasons why participants did not solve the issue after finding it are asked in a subsequent question (see Table \ref{table:impediments_prevented_TAI}).

When participants addressed any issues found, they declared the most implemented strategies were "\textit{improving the quality of the dataset (\textit{e.g.}, removing spurious samples, paradoxical values)}" (8 answers) and "\textit{augmenting the dataset (\textit{e.g.}, with artiﬁcial, manually generated data points)}" (6 answers). On the other hand, the less implemented strategy was "\textit{searching for a tool which automates a specific trustworthiness issue-fixing process}" (1 answer).
One interviewee also mentioned that they usually approach explainability by "\textit{using [only] white box models}". More details can be found in Figure A2 in the online appendix \cite{online_appendix}.

Regarding the reasons why participants 
did not solve a TAI issue after finding it, Table \ref{table:impediments_prevented_TAI} shows the most frequent reasons are "\textit{the issue solution required too much time to be implemented}" (58.3\%) and "\textit{the issue solution was likely to decrease the performance of the system (e.g., decreasing accuracy)}" (50\%). On the other hand, none of the participants answered: "\textit{no one had an idea on how to solve the issue}"; this is a positive result since demonstrates practitioners are conscious of untrustworthiness problems and can formulate hypotheses on how to address them. During the interviews, one participant also mentioned "\textit{data availability}" as an impediment.

\begin{table}[]
\small
\caption{Reasons which prevented participants from addressing/ﬁxing AI trustworthiness issues. 
}
\label{table:impediments_prevented_TAI}
\begin{tabular}{lccc}
\hline
\textbf{Impediment}                                                                                                                               & \textbf{Yes} & \textbf{No} & \textbf{N/A} \\ \hline
No one had idea on how to solve the issue                                                                                                         & 0            & 9           & 3            \\ \hline
\begin{tabular}[c]{@{}l@{}}The issue solution required high human effort, \\ which we could not afford\end{tabular}                               & 5            & 5           & 2            \\ \hline
\begin{tabular}[c]{@{}l@{}}The issue solution was too expensive \\ (ﬁnancially) to address\end{tabular}                                           & 4            & 5           & 3            \\ \hline
\begin{tabular}[c]{@{}l@{}}The issue solution required too much time to\\  be implemented\end{tabular}                                            & 7            & 3           & 2            \\ \hline
\begin{tabular}[c]{@{}l@{}}The issue solution was likely to decrease the\\  performance of the system \\ (\textit{e.g.}, decreasing accuracy)\end{tabular} & 6            & 3           & 3            \\ \hline
\begin{tabular}[c]{@{}l@{}}There was not a tool which automated the \\ fixing process\end{tabular}                                                & 5            & 5           & 2           \\ \hline
\end{tabular}
\end{table}





\begin{figure}[ht]
\includegraphics[scale=0.9]{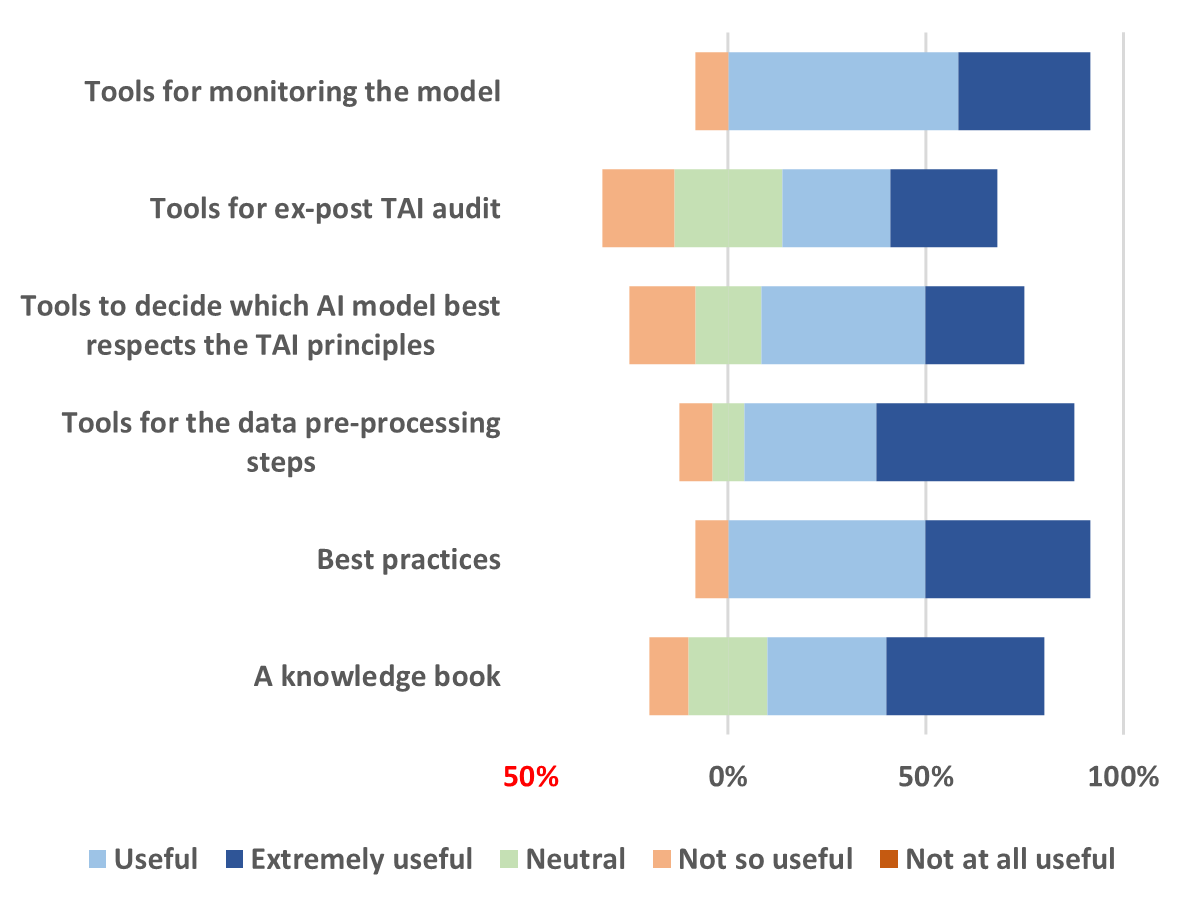}
\caption{Perceived usefulness of hypothetical tools to address TAI issues. N/A answers have been removed.}
\label{fig:tool_address_tai_issues}
\end{figure}

Finally, when we again asked the participants to rate the utility of various hypothetical tools --- assuming their team had access to them --- the participants rated as the most valuable a tool able to "\textit{[...] help [...] monitoring the AI model after its release to the public}", followed by "\textit{best practices that can actively guide your team through the model's SDLC}", "\textit{tools to help the team in the data pre-processing steps (e.g., decide whether one needs to add/remove data points from your training set, and what kind of data you need to add/remove)}", and "\textit{a knowledge book in which are mapped trustworthiness problems and [...] solutions}". On the other hand, they rated as least useful tools "\textit{[...] to help your team doing an ex-post TAI audit}" and tools able to \textit{[...] help your team deciding which AI model best respects the TAI principles [...]}. These results are graphically shown in Fig. \ref{fig:tool_address_tai_issues}.



\subsection{Demographics and background information}\label{sec:demographics}

\textcolor{black}{In terms of demographics of the sample, the study participants’ gender is represented by 68\% male, 21\% female, 3\% non-binary or gender diverse, and 9\% preferred not to respond.}
Concerning academic qualifications, 56\% of the participants have a master's degree, 41\% have earned a Ph.D., and 3\% have completed their bachelor's studies in computer science-related fields. The notable prevalence of advanced education within the respondent pool aligns with our expectations, reflecting the elevated cognitive expertise required by the complexities of this particular field.

The vast majority of the participants are employed in medium-large companies, specifically, 32.3\% (11/34) work for companies with less than 50 employees (small), 23.5\%, (8/34) for companies between 50 and 500 employees (medium), while 44.1\%, (15/34) are employed in large companies with more than 500 employees.


Participants have on average five years of experience in their role and two years of experience in the AI field.

Regarding the technology area that best describes with which AI products/services the participants work, the four most prevalent are "\textit{Decision support}" (15 answers), "\textit{Natural Language Processing}" (13 answers), "\textit{Computer Vision / Image Analysis}" (12 answers), and "\textit{Recommender Systems}" (9 answers). 


Finally, regarding the number of AI-enabled projects developed and deployed into a production environment, we observed that most of the participants (19/34, 56\%) declared that just a small percentage of the developed projects --- from 1 to 30\% --- are deployed in a production environment, while only 3\% of the participants declared that most of the projects --- from 90 to 100\% --- are deployed into a production environment. This reveals the fact that most of these types of projects are still in an experimental stage.

\textcolor{black}{Due to space constraints, we have not included tabular representation of demographics in the paper which are, however, all available in the online appendix \cite{online_appendix}}\footnote{Appendix Table A4 (interviewees’ self-reported technology areas and team roles); Table A5 (participants count grouped by company size); Table A6 (participants count grouped by years of experience in their current role and in developing AI-enabled systems)}.

\subsection{Summary of key findings}
Here we summarize some key findings from our study.

\textbf{F1}. 
\textcolor{black}{The study reveals that participants care a lot about \textbf{Privacy} and \textbf{Transparency}}.  Indeed, among the most used strategies to ensure trustworthiness are "\textit{post-hoc explainability}" and "\textit{algorithm that can best explain the decision}" (Fig. \ref{fig:strategy_prevent_tai_issues}). In addition, tools that "\textit{generate an explanation}" and that help in deciding "\textit{the more clear and explainable model}" are among the tools perceived as most useful (Fig. \ref{fig:tool_prevent_tai_issues}).

\textbf{F2}. Acting on the \textbf{dataset} is one of the most used strategies to solve the found TAI issues. Indeed the most implemented strategies are "\textit{improving the quality of the dataset (e.g., removing spurious samples, paradoxical values)}" and "\textit{augmenting the dataset (e.g., with artiﬁcial, manually generated data points)}".

\textbf{F3}. Business constraints --- like the time required to implement the solution or the unacceptable performance drop --- often represent \textbf{impediments} to implementing trustworthy AI applications (see Table \ref{table:impediments_prevented_TAI}).
    
\textbf{F4}. Analyzing the \textbf{tools} practitioners lack the most in addressing TAI issues, a need for tools for after-deploy \textbf{monitoring} and \textbf{best practices} and TAI \textbf{knowledge books} that can actively guide a team through the SDLC emerges (Fig. \ref{fig:tool_address_tai_issues}).

\textbf{F5}. Many TAI projects are developed but not deployed in a production environment, which reveals that in some cases practitioners are still experimenting with this field.
\section{Discussion}
\label{sec:discussion}

\textbf{RQ1. What vision/opinion do practitioners have about untrustworthiness problems and how often do they address them?}

Our findings reveal that the most addressed principle is \textbf{Privacy}, probably because it is contained in various regulations that exist and must necessarily be complied with (\textit{e.g.}, in Europe the GDPR \cite{EuropeanParliament2016a}).
\textbf{Transparency} is also often taken into consideration, probably because there are domains where it is a fundamental and unavoidable feature required by the law, such as in Healthcare and Financial Services.
On the other hand, the one less addressed is \textbf{Fairness}, perhaps because there are still no clear and shared regulations for this dimension of trustworthiness and everything is left to the initiative and ethical values of those implementing these systems. 
As a result, even large and well-established companies in the industry are often caught up in scandals that damage their reputation and show how even the most popular and widely used algorithms suffer from unfairness\footnote{\url{https://www.bloomberg.com/graphics/2023-generative-ai-bias/}}.

It is notable that when we asked practitioners \textit{why their team cares about trustworthiness in AI}, the motivation "\textit{retain users/avoid losing business}" found most disagreement among them. This may reveal that they believe that TAI issues do not lead to losing users and/or business. Moreover, during interviews, it emerged that addressing TAI is, in some cases, even mandatory and not an option.

Finally, our study suggests that TAI is mainly addressed in the early stages of the SDLC. While this is good --- since the earlier certain decisions are made, the more effective they are in the design of the final model --- this also reveals that practitioners are most likely not aware of tools and best practices to be used in the final stages of the lifecycle. In fact, \textbf{Deploy} is one of the least addressed phases. Indeed, one interviewee also mentioned the need for guidelines on the best cloud provider compliant with TAI practices. These elements allow us to infer that the choice of deployment infrastructure is often left to chance or, in the best case, to routines and/or trust in a specific cloud provider.

\textbf{RQ2. What do practitioners do to address untrustworthiness problems?}

Based on the answers, it is clear that when practitioners want to solve TAI issues they mainly act on the dataset, as the most implemented strategies are related to "improving the quality of the dataset" and "augmenting it with artificial data points".
As shown by Fig. \ref{fig:tai_issues_addressed_SDLC} and Fig. \ref{fig:strategies_discover_tai_issues}, TAI is mainly addressed in \textbf{Design} (\textit{e.g.}, examining model's input features) and \textbf{Development} phase, without disregarding the \textbf{Monitoring} one (\textit{e.g.}, learning from user feedback).
Answers to question 22\footnote{Q22: Which of the following strategies has your team evaluated, and which strategies were actually implemented?} sustain this trend and highlight that practitioners also give little consideration to the possibility of searching for tools that automate their manual activities, perhaps because they feel safer with manual analyses or because they are not familiar with automatic tools.
Finally, answers to question 24\footnote{Q24: What prevented your team from addressing/ﬁxing these AI trustworthiness issues?} demonstrate that business constraints (\textit{e.g.}, time, money) often hinder the resolution of TAI issues: if solving a problem takes too long or may cause a slight performance degradation, practitioners tend to not address it.

\textbf{RQ3. What do practitioners desire to better address untrustworthiness problems?}

What emerged from the data is that the tools most in demand are those that can generate an explanation of the model after its training. 
However, literature provides several mature models to explain both traditional ML algorithms --- see SHAP \cite{NIPS2017_7062} and LIME \cite{lime} --- and neural networks --- such as \cite{KENNY2021103459}. 
Perhaps these answers could be due either to the unsuitability of such tools for specific issues to be solved or to scarce knowledge about explainability tools landscape, which would also explain the fact that \textit{post-hoc explainability} strategies are infrequently used (see Fig. \ref{fig:strategy_prevent_tai_issues}).
On the contrary, the tools deemed less useful are the ones that help decide how much data they need for particular subgroups/subpopulations, probably because there are other factors --- not related to TAI requirements --- that constrain practitioners while collecting and/or pre-processing the data. For instance, sometimes, collecting more data about a specific sub-group could be simply infeasible or very difficult.

Furthermore, it is evident that practitioners feel the need for tools to monitor the model after it has been released to the public, they especially express a need for guidelines and a knowledge base to help them in implementing TAI throughout the SDLC. 
Moreover, practitioners believe a single tool \textit{to perform post-hoc analyses} to be of little use, probably because they feel that it is too late to worry about TAI once the system has been released on the market to end users.


\subsection{Practical implications and recommendations}
Based on the previous discussion, we summarize some key practical implications and recommendations for the AI industry and the AI research community.

\textbf{P1.} Practitioners should take TAI principles into account throughout the entire SDLC and not just in the early stages, such as the \textbf{Design} phase, as shown by Fig. \ref{fig:tai_issues_addressed_SDLC} and answers to question 18\textcolor{black}{\footnote{Q18: Please indicate if you use one of the following strategies to discover trustworthiness issues.}}. Even if valuable mitigation measures can be put in place early in SDLC, there is a huge amount of mitigations that can only be implemented in the latter stages of the SDLC. For instance, using a tool that helps to choose the best cloud provider to be TAI-compliant.

    
\textbf{P2.} Our study reveals that, on numerous occasions, even when TAI issues are identified, they are not addressed by practitioners --- either directly or by third parties --- due to business constraints such as limited time, financial constraints, or declining performance. Nevertheless, the oversight of these issues may result in significant economic and reputational consequences, incurring substantial costs for companies. Hence, it is imperative for the industry to commit to addressing TAI issues detected at various stages of the SDLC. In addition, such actions are crucial for compliance with emerging regulations, such as the AI Act \cite{eu_ai_act}.


\textbf{P3.} As a general remark, there is a pressing need for \textbf{guidelines}, \textbf{knowledge bases}, and \textbf{tools} that can help practitioners implement TAI principles throughout the entire SDLC. They need guidance and practical advice on which tools to use at each stage of SDLC and to address which principles. Often, as pointed out in our study, although these tools exist, practitioners may not be aware of them. Moreover, as pointed out in our previous work \cite{10.1145/3593434.3593478}, there is a significant gap that should be filled between high-level AI ethics principles and low-level concrete practices for practitioners.
For this reason, as a research community, we should rethink how to design these guidelines and best practices, so that they are readily available and usable by professionals and provide actionable guidelines that can be put into practice while implementing trustworthy AI applications.

\section{Threats to validity}
\label{sec:threats}

In this section, we discuss the threats to the validity \cite{Wohlin2012ExperimentationIS} of our study.
We delineate the threats to validity and constraints on the outcomes of our study arising from the research methodology we employed. 

\textit{Construct validity}
The thematic analysis was executed by two researchers, introducing the potential for subjective judgment. To address this concern, we implemented the negotiated agreement technique \cite{10.1145/3180155.3180192} between the first and second researcher, fostering consensus, which was achieved after a careful examination of 25 comments. 

\textit{Internal validity}. Threats to internal validity pertain to unconsidered factors that might impact the variables and relationships under scrutiny. In our investigation, we conducted interviews with AI practitioners to gain insights into their perspectives on Trustworthy AI (TAI) issues. Each practitioner possesses a distinct background and ethical standpoint, potentially diverging from the practices of their peers. We sought to mitigate this issue by interviewing practitioners from diverse companies and different countries. Furthermore, interviewees' viewpoints may be influenced by additional factors, such as existing literature on TAI, potentially leading to social desirability bias \cite{Furnham1986ResponseBS}, or practices adopted in company projects they are involved in. To counteract this, we consistently reminded interviewees that the discussion focused specifically on the TAI issues they encounter in their daily work. At the conclusion of the interviews, we encouraged them to freely express their broader thoughts on TAI.

\textit{Generalizability – Transferability}. One notable threat lies in the generalizability of findings, as our sample, albeit diverse, may not fully represent the broader population of professionals engaged in AI development. To mitigate this threat, we solicited opinions from a heterogeneous participants sample: we took practitioners from corporates, characterized by different sizes 
with different years of experience --- both in their current role and in AI-enabled systems development;
with different genders, level of education, and job roles; working in diverse application domains and technology areas. All details can be found in the online appendix \cite{online_appendix}.
\section{Conclusion}
\label{sec:conclusions}

In this study, we conducted think-aloud \textbf{interviews} and a \textbf{survey} with \textbf{34} AI practitioners to explore how they handle and implement Trustworthy AI applications. 
The study highlights how, among the TAI principles considered, practitioners mainly focus on \textbf{Privacy} at the expense of \textbf{Fairness}, even if many practitioners acknowledge to be aware of how important this last dimension is.
Noteworthy is the fact that half of the participants stated that they did not fix TAI issues after discovering them in their projects. Indeed only half of them declared to have addressed the issue, either by themselves (35\%) or revolving to a third party (15\%).
From the study it also emerged that TAI is mainly addressed in the initial phases of the SDLC (mainly during the \textbf{Design} and \textbf{Development}) and few practitioners declared to also address it in later stages, like \textbf{Deploy}. 
The strategies most employed to build TAI applications focus on data quality enhancement or choosing the most self-explanatory algorithm. Moreover, data and algorithm design are conducted by exploiting manual analysis, without using any automated tool. As one could expect, the most common cause of obstruction in implementing TAI is business constraints (\textit{e.g.}, time, money). 

Finally, we also identified that practitioners feel the need for tools to \textbf{monitor} the model after production deployment, and 
\textbf{knowledge bases} and \textbf{actionable guideline}s to help them implement trustworthiness throughout the entire SDLC.
While different frameworks, tools, and guidelines may exist \cite{10.1145/3593434.3593478}, these are either composed of high-level statements that are sometimes difficult to translate into concrete implementation strategies or may be unknown to AI practitioners (as shown by all the answers to the questions related to the need/usefulness of tools to address TAI issue). As future directions, we intend to complement and expand this work with more interviews with a larger sample of AI practitioners to investigate some of the critical points that emerged in this study more in-depth.


\balance

\begin{acks}
This study has been partially supported by the following projects: SSA (Secure Safe Apulia - Regional Security Center, Codice Progetto 6ESURE5) and KEIRETSU (Codice Progetto V9UFIL5) funded by "Regolamento regionale della Puglia per gli aiuti in esenzione n. 17 del 30/09/2014; SERICS (PE00000014) under the MUR National Recovery and Resilience Plan funded by the European Union – NextGenerationEU.
\end{acks}

\bibliographystyle{ACM-Reference-Format}
\bibliography{sample-base}

\end{document}